# Total/dual correlation/coherence, redundancy/synergy, complexity, and O-information for real and complex valued multivariate data

Roberto D. Pascual-Marqui[1], Kieko Kochi[1], Toshihiko Kinoshita[2]

1: The KEY Institute for Brain-Mind Research; Department of Psychiatry, Psychotherapy, and Psychosomatics; University of Zurich, Switzerland
2: Department of Neuropsychiatry, Kansai Medical University, Osaka, Japan

Corresponding author: RD Pascual-Marqui
robertod.pascual-marqui@uzh.ch ; https://www.uzh.ch/keyinst
https://scholar.google.com/citations?user=DDqjOkUAAAAJ

## 1. Abstract

Firstly, assuming Gaussianity, equations for the following information theory measures are presented: total correlation/coherence (TC), dual total correlation/coherence (DTC), O-information, TSE complexity, and redundancy-synergy index (RSI). Since these measures are functions of the covariance matrix "S" and its inverse "S^-1", the associated Wishart and inverse-Wishart distributions are of note. DTC is shown to be the Kullback-Leibler (KL) divergence for the inverse-Wishart pair "(S^-1)" and its diagonal matrix "D=diag(S^-1)", shedding light on its interpretation as a measure of "total partial correlation", -lndetP, with test hypothesis H0: P=I, where "P" is the standardized inverse covariance (i.e. P=(D^-1/2)(S^-1)(D^-1/2). The second aim of this paper introduces a generalization of all these measures for structured groups of variables. For instance, consider three or more groups, each consisting of three or more variables, with predominant redundancy within each group, but with synergistic interactions between groups. O-information will miss the between group synergy (since redundancy occurs more often in the system). In contrast, the structured O-information measure presented here will correctly report predominant synergy between groups. This is a relevant generalization towards structured multivariate information measures. A third aim is the presentation of a framework for quantifying the contribution of "connections" between variables, to the system's TC, DTC, O-information, and TSE complexity. A fourth aim is to present a generalization of the redundancy-synergy index for quantifying the contribution of a group of variables to the system's redundancy-synergy balance. Finally, it is shown that the expressions derived here directly apply to data from several other elliptical distributions. All program codes, data files, and executables are available (https://osf.io/jd37g/).

## 2. Introduction

Brain function research is benefiting from many recent studies applying information theory concepts and analyses, especially in elucidating integration/segregation of function. This paper mainly focuses on applying a subset of measures from information theory to the analysis of time series data of brain activity. This can be the case of signals of cortical electric neuronal activity, as directly measured with invasive recordings, or estimated as with low resolution electromagnetic tomography (LORETA; Pascual-Marqui 2002 and 2007) from non-invasive EEG or MEG recordings. This is also the case of signals of metabolic activity obtained from PET and fMRI.

Except for some few publications, the description of information theory measures such as redundancy, synergy, O-information, and complexity is systematically given in terms of symbolic "H" and "I" for Shannon entropy and mutual information, for a generic probability distribution, often times not specified. One aim of our work is to present these equations for the multivariate Gaussian probability distribution, in terms that are intelligible and meaningful to those that are familiar with multivariate statistics (e.g. as in Mardia et al 1979) and time series analysis (e.g. as in Brillinger 2001). These statistic-type equations, rather than detract from the formalism of information theory, enrich it, add insight, provide new interpretations, and allow important generalizations not previously found in current information theory publications (to the best of our knowledge).

Excellent sources of practical and applied information theory methods that include some basic equations for real continuous multivariate Gaussian data can be found in Varley et al (2023), Reva et al (2024), and Pope et al (2025).

Other work, related but not equivalent to the methods presented here, can be found in Antonacci et al (2025), Mijatovic et al (2025), and Sparacino et al (2025). Although also related, our present work excludes the methods that specialize in the partial information decomposition of Williams and Beer (2010), which studies the interactions of two predictive vectors and one response vector in terms of synergy, redundancy, and uniquenesses.

## 3. Some notation

$N_{\mathbb{C}}^p(\mathbf{0}, \mathbf{S})$ denotes the multivariate complex Gaussian distribution, with mean zero and positive definite Hermitian covariance matrix $\mathbf{S} \in \mathbb{C}^{p \times p}$. $N_{\mathbb{R}}^p(\mathbf{0}, \mathbf{S})$ denotes the real-valued case, with real symmetric positive definite covariance matrix $\mathbf{S} \in \mathbb{R}^{p \times p}$.

Typically, for all the information measures considered here, the minimum dimension is $p_{\min} = 3$, i.e. beyond the simple bivariate system with just one pairwise interaction.

ln: natural logarithm

det: determinant of a matrix

*diag*: operator that returns a matrix with its off-diagonal elements equal to zero

**I**: identity matrix of appropriate dimension defined by the equation where it appears

*std*: operator that standardizes a positive definite Hermitian matrix **H**:

Eq. 1 $\quad std\mathbf{H} = (diag\mathbf{H})^{-1/2} (\mathbf{H}) (diag\mathbf{H})^{-1/2}$

This operator will later be applied to covariance matrices and to precision/concentration matrices (i.e. inverse of the covariance matrices). Typically, the following symbols are used in this work:
**R**: reserved for correlation and coherence matrix, i.e. the standardized covariance matrix;
**P**: reserved for related partial correlation and partial coherence matrix, i.e. the standardized precision matrix.

For $N^p(\mathbf{0},\mathbf{S})$ data, consider a sample of size "$n$" and let $\hat{\mathbf{S}}$ denote the estimated covariance. Then $n\hat{\mathbf{S}} \sim W^p(n,\mathbf{S})$ has a Wishart distribution, and $(n\hat{\mathbf{S}})^{-1} \sim invW^p(n,\mathbf{S}^{-1})$ has an inverse-Wishart distribution. See e.g. Brillinger (2001) equations 4.2.7 through 4.2.9 therein; Gupta and Srivastava (2010) equations 1 and 2 therein; Mardia et al (1979) equations 3.8.1 and 3.8.2 therein.

Let $\mathbf{X} \in \mathbb{C}^{p \times 1}$ denote a generic complex-valued vector with dimension "$p$". Its Hermitian covariance matrix is $\mathbf{S}_{xx} \in \mathbb{C}^{p \times p}$:

**Eq. 2** $\quad \mathbf{S}_{xx} = \text{var}(\mathbf{X}) = \text{cov}(\mathbf{X},\mathbf{X}) = E(\mathbf{XX}^*)$

where "$E$" denotes the statistical expectation operator corresponding to the average, and the superscript "*" denotes complex conjugate and vector/matrix transpose. In general:

**Eq. 3** $\quad \mathbf{S}_{xy} = \text{cov}(\mathbf{X},\mathbf{Y}) = E(\mathbf{XY}^*)$

$x_i$ denotes the i-th element of $\mathbf{X}$, and $s_{ij}$ an element of $\mathbf{S}$.

**Eq. 4** $\quad \left\{ \mathbf{X}_{\delta(i)} \in \mathbb{C}^{(p-1) \times 1} \text{ or } \mathbb{R}^{(p-1) \times 1} \text{ denotes the vector without the } i-\text{th variable, i.e. without } x_i \right\}$

In time series analysis for stationary processes (see e.g. Brillinger, 2001), the cross-spectral matrices ($\mathbf{S}$ as a function of frequency) are computed from the discrete Fourier transform of $\mathbf{X}$. Explicit detailed equations and further references can be found e.g. in Pascual-Marqui et al (2023), see equations 1 and 2 therein.

An alternative parametric estimator for cross-spectral matrices, which is asymptotically equivalent to the non-parametric periodogram estimator, derives from the multivariate autoregression model, see e.g. Pascual-Marqui et al (2014), equations 1 through 4 therein.

For real-valued data with $\mathbf{X} \in \mathbb{R}^{p \times 1}$, its symmetric covariance matrix is $\mathbf{S}_{xx} \in \mathbb{R}^{p \times p}$. An example for BOLD time series can be found in Puxeddu (2025), see the section "Covariance matrices definition" therein.

For any two generic vectors $\mathbf{X} \in \mathbb{C}^{p \times 1}$ and $\mathbf{Y} \in \mathbb{C}^{q \times 1}$, the conditional variance of $\mathbf{X}$ on $\mathbf{Y}$ is:

**Eq. 5** $\quad \mathbf{S}_{x|y} = \text{var}(\mathbf{X}|\mathbf{Y}) = \mathbf{S}_{xx} - \mathbf{S}_{xy}\mathbf{S}_{yy}^{-1}\mathbf{S}_{yx}$

Shannon entropy for $N_\mathbb{C}^p(\mathbf{0},\mathbf{S})$:

**Eq. 6** $\quad H(\mathbf{X}) = p + p \ln \pi + \ln \det \mathbf{S}$

See e.g. Neeser and Massey (1993), theorem 2 therein. For real-valued data, simply include a factor of ½ in the right-hand side of Eq. 6.

## 4. Total correlation, total coherence (TC)

### 4.A. Definition

As reviewed in Rosas et al (2019), total correlation was defined by Watanabe (1960) as:

**Eq. 7** $$TC(\mathbf{X}) = \sum_{i=1}^{p} H(x_i) - H(\mathbf{X})$$

For $N_{\mathbb{C}}^{p}(\mathbf{0},\mathbf{S})$ data, by way of Eq. 6, it is:

**Eq. 8** $$TC_{\mathbb{C}} = -\ln \det \mathbf{R} = -\ln \det std\mathbf{S} = -\ln \det \mathbf{S} + \ln \det diag\mathbf{S}$$

where **R** is the coherence matrix:

**Eq. 9** $$\mathbf{R} \equiv std\mathbf{S} = (diag\mathbf{S})^{-1/2} (\mathbf{S}) (diag\mathbf{S})^{-1/2}$$

For real-valued data, **R** is known as the correlation matrix, and Eq. 8 includes the factor ½ for the right hand sides:

**Eq. 10** $$TC_{\mathbb{R}} = -\frac{1}{2}\ln \det std\mathbf{S} = -\frac{1}{2}\ln \det \mathbf{R} = \frac{1}{2}(-\ln \det \mathbf{S} + \ln \det diag\mathbf{S})$$

### 4.B. Statistics

The total coherence (TC) given by Eq. 8 is identical, except for a scale factor, to the log-likelihood ratio statistic for total/global independence between all variables, for the null hypothesis H$_0$: **R=I** (see e.g. Mardia et al, 1979, equation 5.3.15 therein).

Moreover, TC can be equivalently written in terms of a Kullback–Leibler (KL) divergence (see Austin, 2018 and 2020) for the Wishart distribution as follows:

**Eq. 11** $$TC(\mathbf{S}) = D_{kl}\left[W^p(1,\mathbf{S}) \| W^p(1,diag\mathbf{S})\right] = -\ln \det \mathbf{S} + \ln \det diag\mathbf{S}$$

See Gupta and Srivastava (2010) equations 1, 2, and 18 therein, and the text immediately after equation 18 therein.

In general, KL-divergence is equivalent to the log-likelihood ratio between two distributions, see Cover (1999), Abe (1996).

### 4.C. Interpretation

TC measures the extent to which the correlation/coherence matrix deviates from the identity matrix, i.e., it quantifies the total global correlations/coherences between all variables.

Note that in the literature, TC is interpreted as a measure of ***redundancy***; see e.g. Puxeddu et al (2025) and citations therein: "Information theory proposes the total correlation (TC) as a proxy for redundancy".

### 4.D. Regarding the notation for total correlation in information theory literature

A commonly used notation in information theory literature for the real-valued Gaussian total correlation is:

**Eq. 12** $$\left\{ \begin{array}{l} \text{information theory literature:} \\ TC(\mathbf{X}) = \frac{-\ln|\mathbf{\Sigma_X}|}{2} \quad and \quad TC^{\mathcal{N}}(\mathbf{X}) = \frac{-\ln(|\mathbf{\Sigma}|)}{2} \\ \text{where } |\mathbf{\Sigma}| \text{ denotes the determinant of the covariance matrix} \end{array} \right.$$

as e.g. in Varley et al (2023, equation 21 therein), in Puxeddu et al (2025, equation 13 therein), and in Varley et al (2025, equation 3 therein). This is correct only if the data has been standardized to zero mean and unit variance, as was the case in the cited studies, ensuring correct values. However, the notation in Eq. 12 can be confusing because in general it is formally incorrect. The correct unambiguous equation should explicitly be expressed in terms of the correlation matrix as in as Eq. 10 above:

**Eq. 13** $\quad TC(\mathbf{X}) = -\frac{1}{2} \ln \det \mathbf{R}$ , with: $\mathbf{R} \equiv (diag \mathbf{\Sigma})^{-1/2} (\mathbf{\Sigma}) (diag \mathbf{\Sigma})^{-1/2}$

## 5. Dual total correlation, dual total coherence (DTC)

### 5.A. Definition

As reviewed in Rosas et al (2019), dual total correlation was defined by Han (1975) as:

**Eq. 14** $\quad DTC(\mathbf{X}) = H(\mathbf{X}) - \sum_{i=1}^{p} H(x_i | \mathbf{X}_{\mu(i)})$

For $N_{\mathbb{C}}^p(\mathbf{0}, \mathbf{S})$ data, it is (see "Appendix A" below for proof):

**Eq. 15** $\quad DTC_{\mathbb{C}} = -\ln \det \mathbf{P} = -\ln \det std(\mathbf{S}^{-1}) = -\ln \det(\mathbf{S}^{-1}) + \ln \det diag(\mathbf{S}^{-1})$

where **P** is:

**Eq. 16** $\quad \mathbf{P} \equiv std(\mathbf{S}^{-1}) = \left[diag(\mathbf{S}^{-1})\right]^{-1/2} (\mathbf{S}^{-1}) \left[diag(\mathbf{S}^{-1})\right]^{-1/2}$

Note that **P** has been denoted in the literature as the "scaled concentration matrix", and is equal to the negative partial coherence matrix, see e.g. Artner et al (2022).

For real-valued data $N_{\mathbb{R}}^p(\mathbf{0}, \mathbf{S})$, Eq. 15 includes the factor ½ for the right hand sides:

**Eq. 17** $\quad DTC_{\mathbb{R}} = -\frac{1}{2} \ln \det \mathbf{P} = -\frac{1}{2} \ln \det std(\mathbf{S}^{-1}) = \frac{1}{2} \left[-\ln \det(\mathbf{S}^{-1}) + \ln \det diag(\mathbf{S}^{-1})\right]$

### 5.B. Statistics

The dual total coherence (DTC) given by Eq. 15 is identical, except for a scale factor, to the log-likelihood ratio statistic for "*total **partial** independence*" between all variables, for the null hypothesis $H_0$: **P=I**. This follows from the Kullback–Leibler (KL) divergence (see Austin, 2018 and 2020) for the inverse-Wishart distribution:

**Eq. 18** $\quad DTC(\mathbf{S}) = D_{kl} \left[ invW^p(1, \mathbf{S}^{-1}) \| invW^p(n, diag(\mathbf{S}^{-1})) \right] = -\ln \det \mathbf{S}^{-1} + \ln \det diag(\mathbf{S}^{-1})$

See Gupta and Srivastava (2010) equations 1, 2, and 18 therein, and the text immediately after equation 18 therein. As written earlier, in general, KL-divergence is equivalent to the log-likelihood ratio between two distributions, see Cover (1999), Abe (1996).

### 5.C. Interpretation

DTC measures the extent to which the global/total partial correlations/coherences deviate from zero, i.e., it quantifies deviation from "*total **partial** independence*".

This clarifies the interpretation and meaning of DTC for continuous multivariate data, which up to the best of our knowledge, was not reported previously.

From here, for continuous multivariate data, it would be more proper to rename "dual total correlation/coherence (DTC)" to:
**TPC: "total partial-correlation" / "total partial coherence"**

Note that for multivariate data **X**, it holds that:

**Eq. 19** $\quad DTC(\mathbf{X}) \equiv TC(\mathbf{U})\ ;\ with:\ \mathbf{U} = \mathbf{S}_{xx}^{-1}\mathbf{X}\ and\ \mathbf{S}_{uu} = \mathbf{S}_{xx}^{-1}$

which justifies the explicit derivations based on the inverse Wishart distribution.

## 6. O-information

### 6.A. Definition

Rosas et al (2019) proposed the O-information metric as a global system measure "capable of characterizing synergy- and redundancy-dominated systems". In addition, Rosas et al (2019) extensively discuss the meaning of synergy/redundancy in this context.

In general, O-information is defined as:

**Eq. 20** $\quad \Omega(\mathbf{X}) = TC(\mathbf{X}) - DTC(\mathbf{X})$

For complex Gaussian data $N_{\mathbb{C}}^p(\mathbf{0}, \mathbf{S})$, it can be written as:

**Eq. 21** $\quad \Omega_{\mathbb{C}} = -\ln\det\mathbf{R} + \ln\det\mathbf{P}$

with **R** given by Eq. 9, and **P** given by Eq. 16. For real-valued data $N_{\mathbb{R}}^p(\mathbf{0}, \mathbf{S})$, Eq. 21 includes the factor ½ for the right hand side:

**Eq. 22** $\quad \Omega_{\mathbb{R}} = -\frac{1}{2}\ln\det\mathbf{R} + \frac{1}{2}\ln\det\mathbf{P}$

### 6.B. Statistics

From Eq. 21 it follows that O-information has a very simple statistical interpretation: it is the log-likelihood ratio test of one simple hypothesis vs another simple hypothesis, namely:

**Eq. 23** $\quad (H_a : \mathbf{R} = \mathbf{I})\ vs\ (H_b : \mathbf{P} = \mathbf{I})$

### 6.C. Interpretation

The mathematical definition of O-information is accompanied by the following interpretation (Rosas et al 2019):
"If $\Omega(\mathbf{X}) > 0$ we say that the system is redundancy dominated, while if $\Omega(\mathbf{X}) < 0$ we say it is synergy dominated."

Inevitably, due to the non-negativity of TC and DTC in Eq. 20 and Eq. 21, it is non-misleading to at least loosely associate:

**Eq. 24** $\quad (SYNERGY)\ \sim\ (DUAL\ TOTAL\ COHERENCE)$

## 7. TSE complexity: definition, interpretation, and approximation

Tononi, Sporns, and Edelman (1994) defined a measure of complexity (denoted TSE) with the aim of quantifying the balance between local segregation (i.e. local specialization) and global integration. An approximation to the original definition is given by Rosas et al (2019):

Eq. 25 $\quad TSE(\mathbf{X}) \approx TC(\mathbf{X}) + DTC(\mathbf{X})$

For multivariate complex $N_{\mathbb{C}}^p(\mathbf{0},\mathbf{S})$ and real $N_{\mathbb{R}}^p(\mathbf{0},\mathbf{S})$ Gaussian data it is:

Eq. 26 $\quad TSE_{\mathbb{C}} \approx -\ln\det\mathbf{R} - \ln\det\mathbf{P} \quad ; \quad TSE_{\mathbb{R}} \approx -\frac{1}{2}\ln\det\mathbf{R} - \frac{1}{2}\ln\det\mathbf{P}$

## 8. The redundancy-Synergy Index (RSI)

### 8.A. Definition

The Redundancy-Synergy Index (RSI) was defined in Gat and Tishby (1999), see also Chechik et al (2002). Unlike the previous measures above, the RSI is not a global system measure. Informally, it measures the redundancy-synergy level of a multivariate variable **X** with respect to the scalar variable "$y$" <u>not included in **X**</u>. It can be written as (see Rosas et al 2024):

Eq. 27 $\quad RSI(\mathbf{X}; y) = TC(\mathbf{X}) - TC(\mathbf{X}|y)$

For multivariate complex $N_{\mathbb{C}}^p(\mathbf{0},\mathbf{S})$ and real $N_{\mathbb{R}}^p(\mathbf{0},\mathbf{S})$ Gaussian data it is:

Eq. 28 $\quad RSI_{\mathbb{C}}(\mathbf{X}; y) = -\ln\det\mathbf{R}_{XX} + \ln\det\mathbf{R}_{X|y} \quad ; \quad RSI_{\mathbb{R}}(\mathbf{X}; y) = -\frac{1}{2}\ln\det\mathbf{R}_{XX} + \frac{1}{2}\ln\det\mathbf{R}_{X|y}$

with:

Eq. 29 $\quad \mathbf{R}_{XX} \equiv std\mathbf{S}_{XX} \quad ; \quad \mathbf{R}_{X|y} \equiv std\mathbf{S}_{X|y} = std\left(\mathbf{S}_{xx} - s_{yy}^{-1}\mathbf{S}_{xy}\mathbf{S}_{yx}\right)$

and where the "$std$" operator is defined in Eq. 1.

### 8.B. Statistics

From Eq. 28 it follows that the RSI is the log-likelihood ratio test of one simple hypothesis vs another simple hypothesis, namely:

Eq. 30 $\quad (H_a : \mathbf{R}_{xx} = \mathbf{I}) \ vs \ (H_b : \mathbf{R}_{x|y} = \mathbf{I})$

### 8.C. Interpretation

In Gat and Tishby (1999), in Chechik et al (2002), and in e.g. Reva et al (2024), the scalar variable "$y$" is distinct in nature from the variables in **X**. For instance, "$y$" can be a behavioral variable, and the variables in **X** are some form of neuronal activities.

The mathematical definition of RSI is accompanied by the following interpretation according to Gat and Tishby (1999):
- If $RSI(\mathbf{X}; y) < 0$ then there is synergy, in the sense that the variables in **X** jointly provide more information on "$y$" than when considered separately (i.e. independently);
- If $RSI(\mathbf{X}; y) > 0$ then there is redundancy, in the sense that the variables in **X** do not provide independent information about "$y$".

## 9. The redundancy-synergy index as a univariate localizer λRSI

By way of introduction, consider the work of Scagliarini et al (2023), defining the concept of "gradients of O-information". The *first order* gradient of O-information can be interpreted as follows: it quantifies the nature of the redundancy/synergy contribution that a variable makes to the whole system, i.e. it "localizes" the role of each variable within a system. The first order gradient of O-information for $x_i \in \mathbf{X}$ is defined as:

Eq. 31 $\quad \partial_i \Omega(\mathbf{X}) = \Omega(\mathbf{X}) - \Omega(\mathbf{X}_{\delta(i)})$

where $\mathbf{X}_{\delta(i)}$ is obtained by deleting $x_i$ from $\mathbf{X}$ (see the definition in Eq. 4). In simple terms, the gradient compares O-information of the whole system $\mathbf{X}$, to that of a copy of the system with the variable of interest $x_i$ deleted, i.e. $x_i$ is marginalized out, it is *not* conditionalized out.

Motivated by the work of Scagliarini et al (2023), we define the "**RSI localizer**": $\lambda RSI$.

Again, consider a vector $\mathbf{X} \in \mathbb{C}^{p \times 1}$ partitioned into two parts: $x_i$ and $\mathbf{X}_{\delta(i)} \in \mathbb{C}^{(p-1) \times 1}$, where $\mathbf{X}_{\delta(i)}$ denotes the vector $\mathbf{X}$ without the i-th variable, i.e. without $x_i$ (see the definition in Eq. 4). For this system, based on Eq. 28, we have:

Eq. 32 $\quad \lambda RSI_{\mathbb{C}}(\mathbf{X}_{\delta(i)}; x_i) = -\ln \det \mathbf{R}_{\delta(i)} + \ln \det \mathbf{R}_{\delta(i)|i} \quad ; \quad \lambda RSI_{\mathbb{R}}(\mathbf{X}_{\delta(i)}; x_i) = -\frac{1}{2} \ln \det \mathbf{R}_{\delta(i)} + \frac{1}{2} \ln \det \mathbf{R}_{\delta(i)|i}$

with (see Eq. 1, Eq. 2, and Eq. 5):

Eq. 33 $\quad \mathbf{R}_{\delta(i)} \equiv std\,var(\mathbf{X}_{\delta(i)}) \quad ; \quad \mathbf{R}_{\delta(i)|i} \equiv std\,var(\mathbf{X}_{\delta(i)} | x_i)$

Thus, $\lambda RSI(\mathbf{X}_{\delta(i)}; x_i)$ quantifies the nature of the contribution that $x_i$ makes to the system's redundancy-synergy balance, i.e. it "localizes" the role of each variable within a system:

- If $\lambda RSI(\mathbf{X}_{\delta(i)}; x_i) < 0$ then $x_i$ contributes to the synergy of the system.

- If $\lambda RSI(\mathbf{X}_{\delta(i)}; x_i) > 0$ then $x_i$ contributes to the redundancy of the system.

This definition is admittedly trivial, yet here the classical interpretation of RSI is turned on its head: localized-RSI quantifies the role of a single variable, rather than quantifying the nature of the influence of the system on a single variable.

Moreover, although loosely related, localized-RSI is distinct from the first order gradient of O-information. Two major differences are:
1. The gradient requires the evaluation of the DTC; while λRSI does not.
2. The gradient compares information measures of a p-variate system with that of a (p-1)-variate system without correction for sheer differences in vector dimensions, while λRSI compares information measures of two (p-1)-variate systems.

## 10. Structured "between-groups" measures: σTC, σDTC, σO-information, and σTSE

Consider the following situation. There is a large system composed of many variables, and it can be meaningfully partitioned into several non-overlapping groups, each with its own number of variables. The question of interest here is to quantify the "redundancy/synergy between groups" accounting for and independent of "within group redundancy/synergy patterns". For instance, in Puxeddu et al (2025), the goal was to quantify "within module redundancy, between module synergy".

We propose here a formal and well-constructed statistical method for estimating the "between" relations from the point of view of information theory using "structured" measures of TC, DTC, O-information, and TSE.

### 10.A. Partitioning variables into groups

Consider a vector **X**, and a partition into $K$ subvectors:

Eq. 34 $\quad \mathbf{X} \equiv \begin{pmatrix} \mathbf{X}_1 \\ \mathbf{X}_2 \\ \vdots \\ \mathbf{X}_K \end{pmatrix}$

with partitioned covariance:

Eq. 35 $\quad \mathbf{S} = \begin{pmatrix} \mathbf{S}_{11} & \mathbf{S}_{12} & \cdots & \mathbf{S}_{1K} \\ \mathbf{S}_{21} & \mathbf{S}_{22} & \cdots & \mathbf{S}_{2K} \\ \vdots & \vdots & \ddots & \vdots \\ \mathbf{S}_{K1} & \mathbf{S}_{K2} & \cdots & \mathbf{S}_{KK} \end{pmatrix}$

and corresponding block-diagonal covariance:

Eq. 36 $\quad blockdiag\mathbf{S} = \begin{pmatrix} \mathbf{S}_{11} & \mathbf{0} & \cdots & \mathbf{0} \\ \mathbf{0} & \mathbf{S}_{22} & \cdots & \mathbf{0} \\ \vdots & \vdots & \ddots & \vdots \\ \mathbf{0} & \mathbf{0} & \cdots & \mathbf{S}_{KK} \end{pmatrix}$

where the matrices of zeros **0** have appropriate context-defined dimensions.

Note that the partitioning defined in Eq. 34, Eq. 35, and Eq. 36 also applies to the inverse covariance matrix (i.e. the precision or concentration matrix), as follows. Let:

Eq. 37 $\quad \mathbf{C} = \mathbf{S}^{-1}$

then:

Eq. 38 $\quad \mathbf{C} = \mathbf{S}^{-1} = \begin{pmatrix} \mathbf{C}_{11} & \mathbf{C}_{12} & \cdots & \mathbf{C}_{1K} \\ \mathbf{C}_{21} & \mathbf{C}_{22} & \cdots & \mathbf{C}_{2K} \\ \vdots & \vdots & \ddots & \vdots \\ \mathbf{C}_{Kk1} & \mathbf{C}_{K2} & \cdots & \mathbf{C}_{KK} \end{pmatrix}$ ; $blockdiag\mathbf{C} \equiv blockdiag\mathbf{S}^{-1} = \begin{pmatrix} \mathbf{C}_{11} & \mathbf{0} & \cdots & \mathbf{0} \\ \mathbf{0} & \mathbf{C}_{22} & \cdots & \mathbf{0} \\ \vdots & \vdots & \ddots & \vdots \\ \mathbf{0} & \mathbf{0} & \cdots & \mathbf{C}_{KK} \end{pmatrix}$

Typically, for all the *structured* information measures considered here, the minimum number of partitions is $K_{\min} = 3$, i.e. beyond the simple 2-groups interaction.

It is notable that all the results presented below for these new structured information measures are well-defined and valid even when each subvector consists of a single variable.

### 10.B. Structured total correlation/coherence: σTC

#### 10.B.1. Definition based on statistics

The structured total coherence measure denoted "σTC" corresponds, except for a scale factor, to the likelihood ratio statistic for independence between sets of variables:

**Eq. 39** $H_0 : \mathbf{S} = blockdiag\mathbf{S}$

See e.g. Morrison (2005), subsection "Independence of sets of K sets of variables", equation 29 therein; Anderson (2003), Chapter 9, equation 18 therein; and Wilks (1935).

In terms of the Kullback–Leibler (KL) divergence (see Austin, 2018 and 2020) for the Wishart distribution it is:

**Eq. 40** $\sigma TC(\mathbf{S}) = D_{kl}\left[W^p(1,\mathbf{S}) \| W^p(1, blockdiag\mathbf{S})\right] = \ln\det(blockdiag\mathbf{S}) - \ln\det\mathbf{S}$

See Gupta and Srivastava (2010) equations 1, 2, and 18 therein, and the text immediately after equation 18 therein. For real valued Gaussian data, the extreme right-hand side of Eq. 40 includes the factor ½.

#### 10.B.2. Interpretation

The structured total coherence "σTC" in Eq. 40 is a measure of total coherence between groups and is independent of the total coherences within each group. The total coherences within each group can be computed with Eq. 8 if the group consists of at least three variables.

### 10.C. Structured dual total correlation/coherence: σDTC

#### 10.C.1. Definition based on statistics

The structured dual total coherence measure denoted "σDTC" corresponds, except for a scale factor, to the likelihood ratio statistic for "***partial** independence between sets of variables*":

**Eq. 41** $H_0 : \left(\mathbf{S}^{-1}\right) = blockdiag\left(\mathbf{S}^{-1}\right)$

In terms of the Kullback–Leibler (KL) divergence (see Austin, 2018 and 2020) for the inverse-Wishart distribution it is:

**Eq. 42** $\sigma DTC(\mathbf{S}) = D_{kl}\left[invW^p(1,\mathbf{S}^{-1}) \| invW^p(1, blockdiag(\mathbf{S}^{-1}))\right] = \ln\det\left[blockdiag(\mathbf{S}^{-1})\right] - \ln\det(\mathbf{S}^{-1})$

See Gupta and Srivastava (2010) equations 1, 2, and 18 therein, and the text immediately after equation 18 therein. For real valued Gaussian data, the extreme right-hand side of Eq. 42 includes the factor ½.

#### 10.C.2. Interpretation

The structured dual total coherence "σDTC" in Eq. 42 is a measure of total partial coherence between groups and is independent of the total partial coherences within each group. The total partial coherences within each group can be computed with Eq. 15 if the group consists of at least three variables.

### 10.D. Structured O-information σΩ

We define the structured O-information measure as:

**Eq. 43**  $\sigma\Omega(\mathbf{X}) = \sigma TC(\mathbf{X}) - \sigma DTC(\mathbf{X})$

For complex Gaussian data, it can be written as:

**Eq. 44**  $\sigma\Omega_{\mathbb{C}} = \left[\ln\det(blockdiag\,\mathbf{S}) - \ln\det\mathbf{S}\right] - \left[\ln\det\left[blockdiag(\mathbf{S}^{-1})\right] - \ln\det(\mathbf{S}^{-1})\right]$

Real-valued data includes the factor ½ for right hand side of Eq. 44.

It has the following interpretation, already accounting for "within group interactions":
- If $\sigma\Omega(\mathbf{X}) > 0$ then the interactions between groups are dominated by redundancy.
- If $\sigma\Omega(\mathbf{X}) < 0$ then the interactions between groups are dominated by synergy.

Note that O-information within each group can be computed with Eq. 21 if the group consists of at least three variables.

### 10.E. Structured TSE complexity

We define the structured TSE complexity measure as:

**Eq. 45**  $\sigma TSE(\mathbf{X}) = \sigma TC(\mathbf{X}) + \sigma DTC(\mathbf{X})$

For complex Gaussian data, it can be written as:

**Eq. 46**  $\sigma TSE_{\mathbb{C}} = \left[\ln\det(blockdiag\,\mathbf{S}) - \ln\det\mathbf{S}\right] + \left[\ln\det\left[blockdiag(\mathbf{S}^{-1})\right] - \ln\det(\mathbf{S}^{-1})\right]$

Real-valued data includes the factor ½ for right hand side of Eq. 46.

Note that TSE within each group can be computed with Eq. 26 if the group consists of at least three variables.

## 11. The structured redundancy-synergy index σRSI as a group localizer

Consider the partition of a system defined by Eq. 34, Eq. 35, and Eq. 36. Now define $\mathbf{X}_{\beta[k]}$ as the vector $\mathbf{X}$ in Eq. 34 with the k-th subvector $\mathbf{X}_k$ deleted.

The structured redundancy-synergy index is defined simply as:

**Eq. 47**  $\sigma RSI(\mathbf{X}_{\beta[k]}; \mathbf{X}_k) = \sigma TC(\mathbf{X}_{\beta[k]}) - \sigma TC(\mathbf{X}_{\beta[k]} | \mathbf{X}_k)$

where $\sigma TC$ is the structured total coherence in Eq. 40. This gives:

**Eq. 48**  $\sigma RSI_{\mathbb{C}}(\mathbf{X}_{\beta[k]}; \mathbf{X}_k) = \left[\ln\det(blockdiag\,\mathbf{S}_{\beta[k]}) - \ln\det\mathbf{S}_{\beta[k]}\right] - \left[\ln\det(blockdiag\,\mathbf{S}_{\beta[k]|k}) - \ln\det\mathbf{S}_{\beta[k]|k}\right]$

with:

**Eq. 49**  $\mathbf{S}_{\beta[k]} = \mathrm{var}(\mathbf{X}_{\beta[k]})$ ; $\mathbf{S}_{\beta[k]|k} = \mathrm{var}(\mathbf{X}_{\beta[k]} | \mathbf{X}_k)$

Real-valued data includes the factor ½ for right hand side of Eq. 48.

The structured redundancy-synergy index σRSI quantifies the nature of the contribution that the group $\mathbf{X}_k$ makes to the redundancy-synergy balance "between groups", i.e. it "localizes" the role of each group within a system formed by a collection of groups. This measure accounts for and is independent of the within-group associations.

Interpretation:

- If $\sigma RSI_{\mathbb{C}}(\mathbf{X}_{\beta[k]}; \mathbf{X}_k) < 0$ then the group $\mathbf{X}_k$ contributes to the synergy between groups, independent of the within-groups interactions.

- If $\sigma RSI_{\mathbb{C}}(\mathbf{X}_{\beta[k]}; \mathbf{X}_k) > 0$ then the group $\mathbf{X}_k$ contributes to the redundancy between groups, independent of the within-groups interactions.

## 12. Contribution of connections to information measures: "single node to network" connections

*1. The context*: Let the *p*-dimensional variable $\mathbf{X}$ represent a network with *p*-nodes (i.e. *p* scalar variables), with covariance matrix $\mathbf{S}_{xx}$. Partition $\mathbf{X}$ into $x_i$ and $\mathbf{X}_{\delta(i)}$.

*2. The question of relevance*: How do the "*connections* between $x_i$ and $\mathbf{X}_{\delta(i)}$" contribute to the system's redundancy-synergy balance?

To answer this question, define:

Eq. 50 $\quad \mathbf{X} = \begin{pmatrix} x_i \\ \mathbf{X}_{\delta(i)} \end{pmatrix}$ ; $\mathbf{S} = \begin{pmatrix} s_{ii} & \mathbf{S}_{i\delta} \\ \mathbf{S}_{\delta i} & \mathbf{S}_{\delta\delta} \end{pmatrix}$ ; $s_{ii} = \text{var}(x_i)$ ; $\mathbf{S}_{i\delta} = \text{cov}(x_i, \mathbf{X}_{\delta(i)})$ ; $\mathbf{S}_{\delta i} = \mathbf{S}_{i\delta}^*$ ; $\mathbf{S}_{\delta\delta} = \text{var}(\mathbf{X}_{\delta(i)})$

The partitioning defined in Eq. 50 also applies to the inverse covariance matrix (i.e. the precision or concentration matrix) as:

Eq. 51 $\quad \mathbf{C} = \mathbf{S}^{-1} = \begin{pmatrix} c_{ii} & \mathbf{C}_{i\delta} \\ \mathbf{C}_{\delta i} & \mathbf{C}_{\delta,\delta} \end{pmatrix} = \begin{pmatrix} \left(s_{ii} - \mathbf{S}_{i\delta}\mathbf{S}_{\delta\delta}^{-1}\mathbf{S}_{\delta i}\right)^{-1} & -\left(s_{ii} - \mathbf{S}_{i\delta}\mathbf{S}_{\delta\delta}^{-1}\mathbf{S}_{\delta i}\right)^{-1}\mathbf{S}_{i\delta}\mathbf{S}_{\delta\delta}^{-1} \\ -\left(s_{ii} - \mathbf{S}_{i\delta}\mathbf{S}_{\delta\delta}^{-1}\mathbf{S}_{\delta i}\right)^{-1}\mathbf{S}_{\delta\delta}^{-1}\mathbf{S}_{\delta i} & \left(\mathbf{S}_{\delta\delta} - s_{ii}^{-1}\mathbf{S}_{\delta i}\mathbf{S}_{i\delta}\right)^{-1} \end{pmatrix}$

### 12.A. Contribution of "$x_i$ - $\mathbf{X}_{\delta(i)}$" connections to the total coherence ($\pi TC$)

Under the assumptions of independence and of conditional independence between $x_i$ and $\mathbf{X}_{\delta(i)}$, the "disconnected covariance matrix" $\mathbf{S}_i^{disconn}$ is:

Eq. 52 $\quad \mathbf{S}_i^{disconn} = \begin{pmatrix} \text{var}(x_i \mid \mathbf{X}_{\delta(i)}) & \mathbf{0} \\ \mathbf{0} & \text{var}(\mathbf{X}_{\delta(i)} \mid x_i) \end{pmatrix} = \begin{pmatrix} \left(s_{ii} - \mathbf{S}_{i\delta}\mathbf{S}_{\delta\delta}^{-1}\mathbf{S}_{\delta i}\right) & \mathbf{0} \\ \mathbf{0} & \left(\mathbf{S}_{\delta\delta} - s_{ii}^{-1}\mathbf{S}_{\delta i}\mathbf{S}_{i\delta}\right) \end{pmatrix}$

and the corresponding total coherence for the disconnected system is (based on Eq. 11):

Eq. 53 $\quad TC_{\mathbb{C}}^{disconn}(i) = \ln \det diag \mathbf{S}_i^{disconn} - \ln \det \mathbf{S}_i^{disconn}$

Finally, the contribution of the *connections* between $x_i$ and $\mathbf{X}_{\delta(i)}$ to the total coherence of the system is defined as:

Eq. 54 $\quad \pi TC_{\mathbb{C}}(i) = TC_{\mathbb{C}} - TC_{\mathbb{C}}^{disconn}(i)$

i.e. it is the total coherence of the intact system minus that of the disconnected system under independence and conditional independence. This is a non-negative quantity.

Eq. 54 in a more explicit form is equivalent to:

Eq. 55 $\quad \pi TC_{\mathbb{C}}(i) = \ln \det diag \mathbf{S} - \ln \det \mathbf{S} - \left[\ln \det diag\left(\mathbf{S}_{\delta\delta} - s_{ii}^{-1}\mathbf{S}_{\delta i}\mathbf{S}_{i\delta}\right) - \ln \det\left(\mathbf{S}_{\delta\delta} - s_{ii}^{-1}\mathbf{S}_{\delta i}\mathbf{S}_{i\delta}\right)\right]$

Note that if $x_i$ and $\mathbf{X}_{\delta(i)}$ are independent to start with, then the contribution (Eq. 54) is naturally zero.

For real-valued data, simply include a factor of ½ in the right-hand side of Eq. 53 and Eq. 55.

### 12.B. Contribution of "$x_i$ - $X_{\delta(i)}$" connections to the dual total coherence ($\pi$DTC)

Recall that the dual total coherence is based on the concentration matrix (i.e. inverse covariance matrix), via the Kullback-Leibler divergence for the inverse-Wishart distribution (see Eq. 18).

From Eq. 51, and in analogy to Eq. 52, the "disconnected concentration matrix" is:

Eq. 56 $\quad \mathbf{C}_i^{disconn} = \begin{pmatrix} \left(c_{ii} - \mathbf{C}_{i\delta}\mathbf{C}_{\delta\delta}^{-1}\mathbf{C}_{\delta i}\right) & \mathbf{0} \\ \mathbf{0} & \left(\mathbf{C}_{\delta\delta} - c_{ii}^{-1}\mathbf{C}_{\delta i}\mathbf{C}_{i\delta}\right) \end{pmatrix} \equiv \begin{pmatrix} s_{ii}^{-1} & \mathbf{0} \\ \mathbf{0} & \mathbf{S}_{\delta\delta}^{-1} \end{pmatrix}$

and the corresponding dual total coherence for the disconnected system is (based on Eq. 18):

Eq. 57 $\quad DTC_{\mathbb{C}}^{disconn}(i) = \ln\det diag\,\mathbf{C}_i^{disconn} - \ln\det \mathbf{C}_i^{disconn}$

Finally, the contribution of the *connections* between $x_i$ and $\mathbf{X}_{\delta(i)}$ to the dual total coherence of the system is defined as:

Eq. 58 $\quad \pi DTC_{\mathbb{C}}(i) = DTC_{\mathbb{C}} - DTC_{\mathbb{C}}^{disconn}(i)$

i.e. it is the dual total coherence of the intact system minus that of the disconnected system under independence and conditional independence. This is a non-negative quantity.

Eq. 58 in a more explicit form is equivalent to:

Eq. 59 $\quad \pi DTC_{\mathbb{C}}(i) = \ln\det diag(\mathbf{S}^{-1}) - \ln\det(\mathbf{S}^{-1}) - \left[\ln\det diag(\mathbf{S}_{\delta\delta}^{-1}) - \ln\det(\mathbf{S}_{\delta\delta}^{-1})\right]$

Note that if $x_i$ and $\mathbf{X}_{\delta(i)}$ are independent in the first place, then the contribution (Eq. 58) is naturally zero.

For real-valued data, simply include a factor of ½ in the right-hand side of Eq. 57 and Eq. 59.

### 12.C. Contribution of "$x_i$ - $X_{\delta(i)}$" connections to O-information and to TSE complexity

The contribution of the *connections* between $x_i$ and $\mathbf{X}_{\delta(i)}$ to the system's O-information $\pi\Omega_{\mathbb{C}}(i)$ and to the system's TSE complexity $\pi TSE_{\mathbb{C}}(i)$ is:

Eq. 60 $\quad \pi\Omega(i) = \pi TC(i) - \pi DTC(i)$

Eq. 61 $\quad \pi TSE(i) = \pi TC(i) + \pi DTC(i)$

using Eq. 55 and Eq. 59.

# 13. Structured "between-groups" connections: contribution to information measures

*1. The context*: A network consisting of many nodes is partitioned into several non-overlapping groups, as outlined in subsection "Partitioning variables into groups" (see Eq. 34 through Eq. 38).

*2. The question of relevance*: How do the "*connections* between the k-th group and all other groups" contribute to the system's between-group redundancy-synergy balance, regardless of the within-group connections?

To answer this question, define:

**Eq. 62**
$$\begin{cases} \mathbf{X} = \begin{pmatrix} \mathbf{X}_k \\ \mathbf{X}_{\beta[k]} \end{pmatrix} ; \; \mathbf{S} = \begin{pmatrix} \mathbf{S}_{kk} & \mathbf{S}_{k\beta} \\ \mathbf{S}_{\beta k} & \mathbf{S}_{\beta\beta} \end{pmatrix} ; \; blockdiag\mathbf{S} = \begin{pmatrix} \mathbf{S}_{kk} & \mathbf{0} \\ \mathbf{0} & blockdiag\mathbf{S}_{\beta\beta} \end{pmatrix} \\ with: \mathbf{S}_{kk} = \text{var}(\mathbf{X}_k) ; \; \mathbf{S}_{k\beta} = \text{cov}(\mathbf{X}_k, \mathbf{X}_{\beta[k]}) ; \; \mathbf{S}_{\beta k} = \mathbf{S}_{k\beta}^* ; \; \mathbf{S}_{\beta\beta} = \text{var}(\mathbf{X}_{\beta[k]}) \end{cases}$$

The partitioning defined in Eq. 62 also applies to the inverse covariance matrix (i.e. the precision or concentration matrix) as:

**Eq. 63**
$$\begin{cases} \mathbf{C} = \mathbf{S}^{-1} = \begin{pmatrix} \mathbf{C}_{kk} & \mathbf{C}_{k\beta} \\ \mathbf{C}_{\beta k} & \mathbf{C}_{\beta\beta} \end{pmatrix} = \begin{pmatrix} (\mathbf{S}_{kk} - \mathbf{S}_{k\beta}\mathbf{S}_{\beta\beta}^{-1}\mathbf{S}_{\beta k})^{-1} & -(\mathbf{S}_{kk} - \mathbf{S}_{k\beta}\mathbf{S}_{\beta\beta}^{-1}\mathbf{S}_{\beta k})^{-1}\mathbf{S}_{k\beta}\mathbf{S}_{\beta\beta}^{-1} \\ -\mathbf{S}_{\beta\beta}^{-1}\mathbf{S}_{\beta k}(\mathbf{S}_{kk} - \mathbf{S}_{k\beta}\mathbf{S}_{\beta\beta}^{-1}\mathbf{S}_{\beta k})^{-1} & (\mathbf{S}_{\beta\beta} - \mathbf{S}_{\beta k}\mathbf{S}_{kk}^{-1}\mathbf{S}_{k\beta})^{-1} \end{pmatrix} \\ blockdiag\mathbf{C} = \begin{pmatrix} \mathbf{C}_{kk} & \mathbf{0} \\ \mathbf{0} & blockdiag\mathbf{C}_{\beta\beta} \end{pmatrix} = \begin{pmatrix} (\mathbf{S}_{kk} - \mathbf{S}_{k\beta}\mathbf{S}_{\beta\beta}^{-1}\mathbf{S}_{\beta k})^{-1} & \mathbf{0} \\ \mathbf{0} & blockdiag(\mathbf{S}_{\beta\beta} - \mathbf{S}_{\beta k}\mathbf{S}_{kk}^{-1}\mathbf{S}_{k\beta})^{-1} \end{pmatrix} \end{cases}$$

## 13.A. Contribution of "connections between the k-th group and all other groups" to the structured total coherence (κTC)

The "disconnected covariance matrix" $\mathbf{S}_k^{disconn}$, under the assumptions of independence and of conditional independence between $\mathbf{X}_k$ and $\mathbf{X}_{\beta[k]}$ is:

**Eq. 64**
$$\mathbf{S}_k^{disconn} = \begin{pmatrix} \text{var}(\mathbf{X}_k | \mathbf{X}_{\beta[k]}) & \mathbf{0} \\ \mathbf{0} & \text{var}(\mathbf{X}_{\beta[k]} | \mathbf{X}_k) \end{pmatrix} = \begin{pmatrix} (\mathbf{S}_{kk} - \mathbf{S}_{k\beta}\mathbf{S}_{\beta\beta}^{-1}\mathbf{S}_{\beta k}) & \mathbf{0} \\ \mathbf{0} & (\mathbf{S}_{\beta\beta} - \mathbf{S}_{\beta k}\mathbf{S}_{kk}^{-1}\mathbf{S}_{k\beta}) \end{pmatrix}$$

and the corresponding structured total coherence for the disconnected system is (based on Eq. 40):

**Eq. 65** $\quad \sigma TC_{\mathbb{C}}^{disconn}(k) = \ln\det blockdiag\mathbf{S}_k^{disconn} - \ln\det \mathbf{S}_k^{disconn}$

Finally, the contribution of the contribution of "connections between the k-th group and all other groups" to the structured total coherence of the system is defined as:

**Eq. 66** $\quad \kappa TC_{\mathbb{C}}(k) = \sigma TC_{\mathbb{C}} - \sigma TC_{\mathbb{C}}^{disconn}(k)$

i.e. it is the structured total coherence of the intact system minus that of the disconnected system under independence and conditional independence. This is a non-negative quantity.

Eq. 66 in a more explicit form is equivalent to:

**Eq. 67** $\quad \kappa TC_{\mathbb{C}}(k) = \ln\det blockdiag\mathbf{S} - \ln\det \mathbf{S} - \left[\ln\det blockdiag(\mathbf{S}_{\beta\beta} - \mathbf{S}_{\beta k}\mathbf{S}_{kk}^{-1}\mathbf{S}_{k\beta}) - \ln\det(\mathbf{S}_{\beta\beta} - \mathbf{S}_{\beta k}\mathbf{S}_{kk}^{-1}\mathbf{S}_{k\beta})\right]$

Note that if $\mathbf{X}_k$ and $\mathbf{X}_{\beta[k]}$ are independent to start with, then the contribution (Eq. 66) is naturally zero.

For real-valued data, simply include a factor of ½ in the right-hand side of Eq. 65 and Eq. 67.

### 13.B. Contribution of "connections between the k-th group and all other groups" to the structured dual total coherence (κDTC)

Recall that the structured dual total coherence is based on the concentration matrix (i.e. inverse covariance matrix), via the Kullback-Leibler divergence for the inverse-Wishart distribution (see Eq. 42).

From Eq. 63, and in analogy to Eq. 64, the "disconnected concentration matrix" is:

**Eq. 68** $\quad \mathbf{C}_k^{disconn} = \begin{pmatrix} \left(\mathbf{C}_{kk} - \mathbf{C}_{k\beta}\mathbf{C}_{\beta\beta}^{-1}\mathbf{C}_{\beta k}\right) & \mathbf{0} \\ \mathbf{0} & \left(\mathbf{C}_{\beta\beta} - \mathbf{C}_{\beta k}\mathbf{C}_{kk}^{-1}\mathbf{C}_{k\beta}\right) \end{pmatrix} \equiv \begin{pmatrix} \mathbf{S}_{kk}^{-1} & \mathbf{0} \\ \mathbf{0} & \mathbf{S}_{\beta\beta}^{-1} \end{pmatrix}$

and the corresponding structured dual total coherence for the disconnected system is (based on Eq. 42):

**Eq. 69** $\quad \sigma DTC_{\mathbb{C}}^{disconn}(k) = \ln \det blockdiag \mathbf{C}_k^{disconn} - \ln \det \mathbf{C}_k^{disconn}$

Finally, the contribution of "connections between the k-th group and all other groups" to the structured dual total coherence of the system is defined as:

**Eq. 70** $\quad \kappa DTC_{\mathbb{C}}(k) = \sigma DTC_{\mathbb{C}} - \sigma DTC_{\mathbb{C}}^{disconn}(k)$

i.e. it is the structured dual total coherence of the intact system minus that of the disconnected system under independence and conditional independence. This is a non-negative quantity.

Eq. 70 in a more explicit form is equivalent to:

**Eq. 71** $\quad \kappa DTC_{\mathbb{C}}(k) = \ln \det blockdiag(\mathbf{S}^{-1}) - \ln \det(\mathbf{S}^{-1}) - \left[\ln \det blockdiag(\mathbf{S}_{\beta\beta}^{-1}) - \ln \det(\mathbf{S}_{\beta\beta}^{-1})\right]$

Note that if $\mathbf{X}_k$ and $\mathbf{X}_{\beta[k]}$ are independent to start with, then the contribution (Eq. 70) is naturally zero.

For real-valued data, simply include a factor of ½ in the right-hand side of Eq. 69 and Eq. 71.

### 13.C. Contribution of "connections between the k-th group and all other groups" to O-information and to TSE complexity

The contribution of "connections between the k-th group and all other groups" to the system's O-information $\kappa\Omega(k)$ and to the system's TSE complexity $\kappa TSE(k)$ is:

**Eq. 72** $\quad \kappa\Omega(k) = \kappa TC(k) - \kappa DTC(k)$

**Eq. 73** $\quad \kappa TSE(k) = \kappa TC(k) + \kappa DTC(k)$

using Eq. 66 and Eq. 70.

## 14. Matrix trace equations for TC and DTC

In several cases above (see Eq. 8, Eq. 15, Eq. 40, Eq. 42), the total coherence and the dual total coherence have the generic forms:

**Eq. 74** $\quad TC = -\ln \det \mathbf{R} \quad and \quad DTC = -\ln \det \mathbf{P}$

with:

**Eq. 75** $\quad \begin{cases} \det \mathbf{R} \equiv \det\left[(diag\mathbf{S})^{-1}\mathbf{S}\right] \ or \ \det \mathbf{R} \equiv \det\left[(blockdiag\mathbf{S})^{-1}\mathbf{S}\right] \ and \\ \det \mathbf{P} \equiv \det\left[(diag\mathbf{S}^{-1})^{-1}\mathbf{S}^{-1}\right] \ or \ \det \mathbf{P} \equiv \det\left[(blockdiag\mathbf{S}^{-1})^{-1}\mathbf{S}^{-1}\right] \end{cases}$

The following approximations, which are common in statistics, may be used here:

**Eq. 76** $\quad TC \approx \frac{1}{2}tr\left[(\mathbf{I}-\mathbf{R})^2\right] \quad and \quad DTC \approx \frac{1}{2}tr\left[(\mathbf{I}-\mathbf{P})^2\right]$

See e.g. Morrison (2005), page 40, equations 8 and 10 therein. Moreover, the measures in Eq. 76 are alternative statistics for the same likelihood ratio tests for independence between all scalar variables or between all vector variables, known as the trace statistics of Nagao (1973).

Note that these approximations apply only to TC and to DTC based on "coherence or partial coherence type matrices". They do not apply to the logarithm of the determinant of arbitrary positive definite matrices.

The usefulness of these approximations is that they bypass the computation of the matrix determinant, which may be prone to numerical errors for large dimensions. However, matrix inverse calculation is always needed even if it is also prone to numerical errors for large dimensions. There is a vast literature regarding matrix inversion for large dimensions, and this topic is not treated here.

## 15. Toy examples and discussion

The toy examples considered below are generated cross-spectral matrices, from which the information measures will be computed. The model chosen for their generation is the autoregression, which allows the design of clear unambiguous "connections" that can be redundancy or synergy dominated. In real-life, when using measured time series such as cortical electric neuronal activity or BOLD signals, the cross-spectra can be computed non-parametrically as the periodogram, without the need of fitting an autoregressive model.

### 15.A. Granger causal multivariate autoregression (AR) network

A network with 6 nodes is defined here as a stable AR model of order 2:

**Eq. 77** $\quad \mathbf{X}(t) = \mathbf{A}(1)\mathbf{X}(t-1) + \mathbf{A}(2)\mathbf{X}(t-2) + \mathbf{E}(t)$

with $\mathbf{X}(t), \mathbf{E}(t) \in \mathbb{R}^{p \times 1}$ ; $\mathbf{A}(1), \mathbf{A}(2) \in \mathbb{R}^{p \times p}$, $p=6$, "$t$" is discrete time, and $\mathbf{E} \sim N(\mathbf{0},\mathbf{I})$ independent innovations noise. Figure 1 displays the causal directional connections of the network and the AR coefficients.

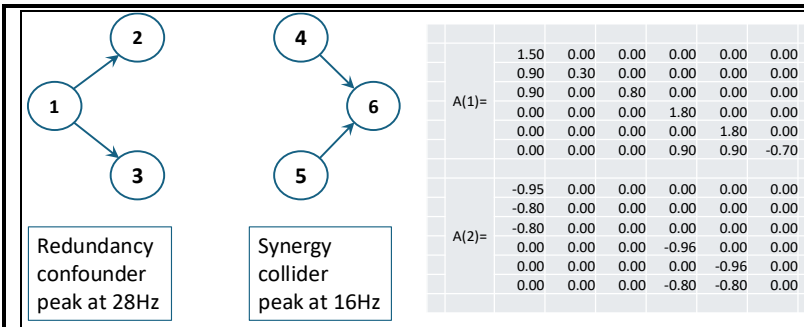

Figure 1: Network with 6 nodes corresponding to a stable AR model of order 2. By design, the first three nodes are in a "redundancy-dominated configuration" and oscillating maximally at 28 Hz; the last three nodes, independent of the first three, are in a "synergy-dominated configuration" and oscillating maximally at 16 Hz. See Rosas et al (2024), figure-1 therein, on redundancy/synergy dominated graphs. Nominal sampling rate is 256 Hz.

By design, the first three nodes are in a "redundancy-dominated configuration" and oscillating maximally at 28 Hz; the last three nodes, independent of the first three, are in a "synergy-dominated configuration" and oscillating maximally at 16 Hz. See Rosas et al (2024), figure-1 therein, on redundancy/synergy dominated graphs. Nominal sampling rate is 256 Hz.

*The question of relevance here*: based on the estimated cross-spectra, can the information measures studied here properly identify redundancy/synergy balance in the frequency domain, globally for the whole network and locally for each node's contribution?

The population Hermitian cross-spectral matrices can be computed directly from the AR coefficients and the innovations covariance matrix ($\mathbf{I}$), as in e.g. Pascual-Marqui et al (2014), equations 1 through 4 therein. However, we prefer a more challenging method, as a stress test: 100 epochs of 256 discrete time-samples each were generated; the nominal sampling rate is 256 Hz, the cross-spectral matrices were estimated as the non-parametric periodogram, via the discrete Fourier transform with a Hann taper, averaging single trial cross-spectra over 100 epochs, as in Pascual-Marqui et al (2023), see equations 1 and 2 therein.

Figure 2 collects all spectra for the information measures studied here.

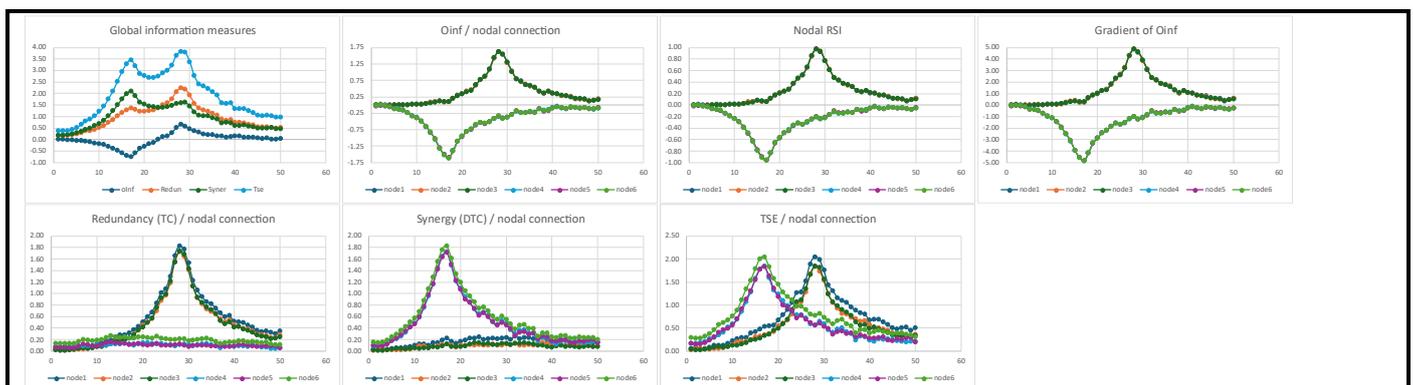

Figure 2:
"Global information measures": O-information Eq. 21; Redun=TC Eq. 8; Syner=DTC Eq. 15; TSE Eq. 26.
"Oinf nodal connection": Eq. 60; "Nodal RSI": Eq. 32. "Gradient of Oinf": Eq. 31.
Redundancy nodal connection Eq. 55; Synergy nodal connection Eq. 59; TSE nodal connection Eq. 61.

From a qualitative point of view all measures plotted in Figure 2 show the expected correct results.

Figure 2: Global information measures: global Oinf is negative (dominant synergy) at 16 Hz and positive (dominant redundancy) at 28 Hz.

Figure 2: Oinf/NodalConnection, Nodal RSI, and Gradient of Oinf: all correctly discriminate the first three nodes as redundancy-contributors with the most positive value at 28 Hz, and the last three nodes as synergy-contributors with the most negative value at 16 Hz.

The added value of the new measures of nodal connection contribution shows near zero redundancy (total coherence) for the last three synergistic nodes (Figure 2: "Redundancy (TC) / nodal connection"), and near zero synergy (dual total coherence) for the first three redundancy nodes (Figure 2: "Synergy (DTC) / nodal connection").

### 15.B. Toy system for groups of nodes: within-group sets redundancy, between-group synergy

In this example a network with 12 nodes is defined here as a stable AR model of order 1:

Eq. 78    $\mathbf{X}(t) = \mathbf{A}(1)\mathbf{X}(t-1) + \mathbf{E}(t)$

with $\mathbf{X}(t), \mathbf{E}(t) \in \mathbb{R}^{p \times 1}$ ; $\mathbf{A}(1) \in \mathbb{R}^{p \times p}$, $p = 12$, "$t$" is discrete time, and each innovation is the sum of an independent $N(0,1)$ Gaussian random variable plus a sine wave of a certain frequency.

The nodes are structured into 4 groups, each group has within-redundancy connections, while there is between-groups synergy, as shown in Figure 3.

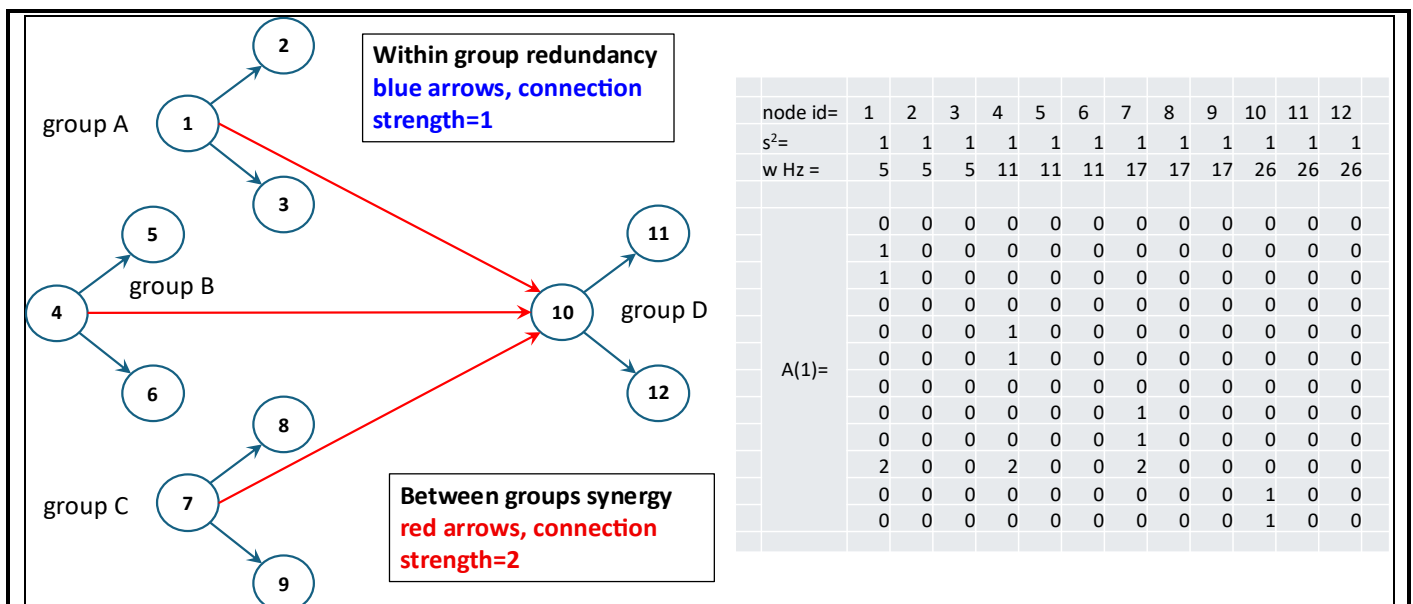

Figure 3: Network with 12 nodes corresponding to a stable AR model of order 1. By design, the nodes are structured into 4 groups, each group has within-redundancy connections, while there is between-groups synergy. All innovations consist of the sum of an independent $N(0,\sigma^2=1)$ Gaussian random variable plus a sine wave with the indicated frequency ($\omega$ Hz). Nominal sampling rate is 256 Hz.

*The question of relevance here*: based on the estimated cross-spectra, can the new information measures developed here properly identify the between-groups synergy, regardless of the larger number of redundancy-dominated connections?

The simulation proceeds as previously described: 100 epochs of 256 discrete time-samples each were generated; the nominal sampling rate is 256 Hz, the cross-spectral matrices were estimated as the non-parametric periodogram, via the discrete Fourier transform with a Hann taper, averaging single trial cross-spectra over 100 epochs, as in Pascual-Marqui et al (2023), see equations 1 and 2 therein.

Figure 4 contains three plots:
**1. Left:** The blue line is the global O-information for all 12 nodes (Eq. 21), showing that the global system is redundancy-dominated (Oinf>0). The red line is the structured O-information (Eq. 43) for 4 groups, which is independent of the within-groups interactions, showing that the system of 4 groups is synergy-dominated (Oinf<0). These are the expected results.
**2. Center:** Contribution to O-information, of the connections between each group and the rest of the system (Eq. 72). As expected, each group contributes synergy (negative values) to the network of groups, independent of the within-group interactions.
**3. Right:** Structured σRSI group localizer (Eq. 47). The structured redundancy-synergy index σRSI as a group localizer. As expected, all values are negative, indicating that each group contributes synergy to the network of groups, independent of the within-group interactions.

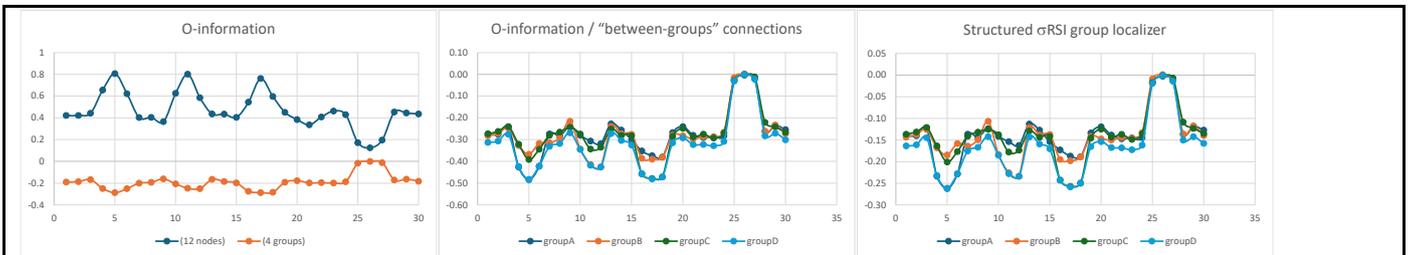

Figure 4:
**1. Left:** The blue line is the global O-information for all 12 nodes (Eq. 21), showing that the global system is redundancy-dominated (Oinf>0). The red line is the structured O-information (Eq. 43) for 4 groups, which is independent of the within-groups interactions, showing that the system of 4 groups is synergy -dominated (Oinf<0).
**2. Center:** Contribution to O-information, of the connections between each group and the rest of the system (Eq. 72). Each group contributes synergy (negative values) to the network of groups, independent of the within-group interactions.
**3. Right:** Structured σRSI group localizer (Eq. 47). The structured redundancy-synergy index σRSI as a group localizer. All values are negative, indicating that each group contributes synergy to the network of groups, independent of the within-group interactions.

## 16. Concluding remarks

**1:** Assuming Gaussianity, equations for the following information theory measures are presented: total correlation/coherence (TC), dual total correlation/coherence (DTC), O-information, TSE complexity, and the redundancy-synergy index (RSI). Since these measures are functions of the covariance matrix $\mathbf{S}$ and its inverse $\mathbf{S}^{-1}$, the associated Wishart and inverse-Wishart distributions are of note.

**2:** The DTC is shown here to be the Kullback-Leibler (KL) divergence for the inverse-Wishart pair "$\left(\mathbf{S}^{-1}\right)$" and its diagonal matrix "$\left(diag\mathbf{S}^{-1}\right)$", shedding light on its interpretation as a measure of "total partial correlation":

$$DTC = -\ln \det \mathbf{P} \text{, with } \mathbf{P} = \left[ diag\left(\mathbf{S}^{-1}\right) \right]^{-1/2} \left(\mathbf{S}^{-1}\right) \left[ diag\left(\mathbf{S}^{-1}\right) \right]^{-1/2}$$

with test hypothesis $H_0 : \mathbf{P} = \mathbf{I}$.

**3:** All information measures treated here are generalized to the case of structured groups of variables. The structured measures quantify the nature of the interactions between groups, accounting for and therefore regardless of the nature of the interactions within each group. The toy example presented here consisted of "within module redundancy, between module synergy". Common global O-information misses the existence of any synergy. In contrast, the structured O-information measure correctly reports predominant synergy between groups. This is a relevant generalization towards structured multivariate information measures.

**4:** A framework is presented for quantifying the contribution of "connections" between nodes, to the system's TC, DTC, O-information, and TSE complexity. This provides the means for "localizing" which connections, if any, contribute to the system's redundancy or synergy, especially since TC (i.e. redundancy) and DTC (i.e. synergy) can be separately and meaningfully estimated.

**5:** The redundancy-synergy index is generalized for quantifying and localizing the contribution of groups of variables to the system's redundancy-synergy balance.

**6:** Important generalizations of the Wishart and inverse Wishart distributions are known as the elliptical Wishart and inverse elliptical Wishart distributions.
**6a:** Ayadi et al (2024) and Hediger et al (2023) show that the estimated dispersion matrices (i.e. the covariance matrix for Gaussian data) are identical, except for a scale factor, to the covariance matrix, with the scale factor depending only on dimension and sample size.
**6b:** Zografos and Nadarajah (2005) show that the entropy of elliptical distributions is identical in form, except for some additive constants (depending only on dimension and sample size), to that of the Gaussian (Eq. 6), with the dispersion matrix in place of the covariance.
**6c:** All this implies that all the information measures studied here, based on relative entropies (i.e. KL-divergencies) for Gaussian data can be applied "as is" to general elliptical data, due to the nature of the KL-divergence, which is invariant to a scale factor for $\mathbf{S}$ and to additive constants.
**6d:** Our results disagree with the recent work of Hindriks et al (2025), where they find differences between Gaussian and the elliptical symmetric Pearson distribution for TC, DTC, and O-information.

**7:** The methods presented here apply to covariance matrices in general. The Hermitian cross-spectral matrices, estimated from multiple time-series, are one interesting example that provide the frequency decomposition of TC, DTC, TSE, O-information, and RSI, in many forms: global, structured by groups (i.e. sets), and the contributions of parts and of connections. In any case, the methods are very general and can be applied to all forms of continuous multivariate data that comply with elliptical distributions.

**8:** It is notable that all information measures for structured data (referring to interactions between sets of nodes) are well-defined and valid even when each set consists of a single variable. In this case, all results are identical to the corresponding non-structured measure.

**9:** Coming up next: TC, DTC, TSE, RSI, and Oinf analysis of real human electrophysiological time series signals of estimated cortical electric neuronal activity, at brain sites corresponding to the major nodes of resting state networks (Uddin et al, 2019).

**10:** All program codes, data files, and executables are available, allowing the interested reader to: test, check, reproduce, and validate. Language: Delphi PASCAL. Executable: Windows OS. https://osf.io/jd37g/

## 18. Appendix A

Here we derive the new equations Eq. 15 and Eq. 16 for dual total coherence (DTC), which we suggest be renamed to total partial coherence (TPC).

From the basic definitions in Eq. 6 and Eq. 14 we have (also see Eq. 4):

Eq. 79 $$DTC(\mathbf{X}) = H(\mathbf{X}) - \sum_{i=1}^{p} H\left(x_i | \mathbf{X}_{\mu(i)}\right) = \left(p + p\ln\pi + \ln\det\mathbf{S}_{xx}\right) - \sum_{i=1}^{p}\left[1 + \ln\pi + \ln\det\mathrm{var}\left(x_i | \mathbf{X}_{\mu(i)}\right)\right]$$

From Erb (2020), equation 9 therein, we have:

Eq. 80 $$\mathrm{var}\left(x_i | \mathbf{X}_{\mu(i)}\right) = \frac{1}{\left[\mathbf{S}_{xx}^{-1}\right]_{ii}}$$

where $\left[\mathbf{S}_{xx}^{-1}\right]_{ii}$ is the i-th diagonal element of $\mathbf{S}_{xx}^{-1}$. This gives:

Eq. 81 $$\sum_{i=1}^{p} \ln\mathrm{var}\left(x_i | \mathbf{X}_{\mu(i)}\right) = -\ln\det diag\left(\mathbf{S}_{xx}^{-1}\right)$$

Plugging Eq. 81 into Eq. 79, step by step, finally proves Eq. 15 and Eq. 16:

Eq. 82
$$\begin{cases}
DTC(\mathbf{X}) = \left(p + p\ln\pi + \ln\det\mathbf{S}_{xx}\right) - p - p\ln\pi - \sum_{i=1}^{p}\ln\det\mathrm{var}\left(x_i | \mathbf{X}_{\mu(i)}\right) \\
= \ln\det\mathbf{S}_{xx} - \sum_{i=1}^{p}\ln\det\mathrm{var}\left(x_i | \mathbf{X}_{\mu(i)}\right) \\
= \ln\det\mathbf{S}_{xx} + \ln\det diag\left(\mathbf{S}_{xx}^{-1}\right) \\
= -\ln\det\mathbf{S}_{xx}^{-1} + \ln\det diag\left(\mathbf{S}_{xx}^{-1}\right) \\
= -\ln\det\left\{\left[diag\left(\mathbf{S}_{xx}^{-1}\right)\right]^{-1/2}\left(\mathbf{S}_{xx}^{-1}\right)\left[diag\left(\mathbf{S}_{xx}^{-1}\right)\right]^{-1/2}\right\} \\
= -\ln\det\mathbf{P}
\end{cases}$$